\documentclass[12pt]{article}

\catcode`\@=11

\global\arraycolsep=2pt
\usepackage{amsbsy,amssymb,latexsym,amsfonts,amsmath}
\usepackage{graphicx,color}

\begin{document}
\begin{flushright}
\parbox{4.2cm}
{RUP-18-6}
\end{flushright}

\vspace*{0.7cm}

\begin{center}
{ \Large Local field theory construction of Very Special Conformal Symmetry}
\vspace*{1.5cm}\\
{Yu Nakayama}
\end{center}
\vspace*{1.0cm}
\begin{center}

Department of Physics, Rikkyo University, Toshima, Tokyo 171-8501, Japan

\vspace{3.8cm}
\end{center}

\begin{abstract}
Cohen and Glashow argued that very special conformal field theories of a particular kind (i.e. with HOM(2) or SIM(2) invariance) cannot be constructed within the framework of local field theories. We, however, show examples of local construction by using non-linear realization. We further construct linear realization from the topological twist at the cost of unitarity. To demonstrate the ubiquity of our idea, we also present corresponding holographic models.
\end{abstract}

\thispagestyle{empty} 

\setcounter{page}{0}

\newpage




\section{Introduction}
The assumption of locality plays an essential role in relativistic quantum field theories. In particle physics, it is usually argued that locality is necessary to guarantee the causal structure that is compatible with the special relativity: ``nothing can travel faster than the speed of light". This, however, implies that the motivation to impose locality comes from the more sacred principle of causality and may not be fundamental. Indeed, with extended objects such as branes or strings, the interaction may take place in a non-local way, but still is compatible with the relativistic causality. In this sense, we may say that locality is tied up with the notion of particles under the assumption of the special relativity.

If we abandon the special relativity, the role of locality becomes less obvious. However, Cohen and Glashow argued that the locality may also play a significant role in very special relativity \cite{Cohen:2006ky}\cite{Cohen:2006ir}, which is a certain subgroup of Lorentz symmetry that preserves a particular null direction. They claim that if they impose the locality in field theories that obey a certain class of the very special relativity, they must be fully Lorentz invariant. The claim yields a direct connection between the violation of Lorentz symmetry and the violation of locality, which makes the very special relativity more predictive and interesting. In addition, the speed of light is constant in every directions even though we have a particular null direction, so such theories are phenomenologically viable.

Their argument was based on the spurion analysis. Suppose we begin with a relativistic field theory and consider its local deformation to break the symmetry down to particular subgroups of the Lorentz symmetry (technically known as SIM(2) or HOM(2) invariant very special relativity to be defined below). Cohen and Glashow found that there are no such local operators available from the representation theory of Lorentz algebra. Therefore, they argue that there are no local field theories that realize SIM(2) and HOM(2) invariant very special relativity without symmetry enhancement to the full Lorentz symmetry. Alternatively, they proposed a way to achieve this by violating the assumption of locality at the same time \cite{Cohen:2006ir}.\footnote{See e.g. an explicit background-field origin of this non-locality in QED \cite{Ilderton:2016rqk}.}

In this paper, we, however, point out that there is a loophole in their argument. When the original theory possesses a further global symmetry, one may construct the deformation that preserves the very special relativity without violating the locality. We show some examples in the context of very special conformal field theories \cite{Nakayama:2017eof} for definiteness, but the similar construction is possible and obviously easier without imposing the conformal symmetry.

\section{Very special conformal symmetry}
To discuss very special relativity as well as very special conformal field theories, it is convenient to introduce light-cone coordinate: $x^+ = \frac{1}{\sqrt{2}}(x^0 + x^1)$, $x^- = \frac{1}{\sqrt{2}}(x^0-x^1)$ and $x^i$ ($i=2,3$) (or $x_+ = -\frac{1}{\sqrt{2}}(x^0-x^1)$, $x_- = -\frac{1}{\sqrt{2}}(x^0+x^1)$ and $x_i = x^i$). The light-cone tensors are defined in a similar manner.

The very special relativity is based on the algebra spanned by $P_+$, $P_-$, $P_i$ and $J_{+i}$, where $P_\mu = \{ P_+, P_-, P_i \}$ are space-time translation and $J_{+i}$ is a Lorentz transformation that preserves a particular null direction. In \cite{Cohen:2006ky}, they proposed four different algebra of very special relativity, but in this paper, we focus on either the HOM(2) invariant case by adding $J_{+-}$ or mainly on the SIM(2) invariant case by adding $J_{+-}$ and $J_{ij}$. If we abandon $J_{+-}$ in each case, we have E(2) or T(2) invariant very special conformal field theories respectively.\footnote{The total very special relativity algebra has various names in the literature. The combination of $E(2)$ and $P_\mu$ is sometimes called the Bargmann algebra or massive Galilean algebra (see e.g. \cite{Andringa:2010it} and reference therein). The combination of $SIM(2)$ and $P_\mu$ is called $ISIM(2)$ algebra in \cite{Gibbons:2007iu}.}

The conformal extension of the algebra of very special relativity was discussed in \cite{Nakayama:2017eof}. The gist is that we can only add the dilatation $\tilde{D}$ and a particular special conformal transformation $K_+$ (as a subgroup of the Poincar\'e conformal algebra). The schematic form of the commutation relation for the SIM(2) invariant very special conformal algebra is summarized in table 1. For the HOM(2) invariant case, one can just ignore the column and row of $J_{ij}$. The relevant fact that we will use later is that $J_{+-}$ does not appear in the table as a result of the commutator.

\begin{table}[htb]
  \begin{tabular}{|c|cccccccc|} \hline
     & $P_+$ & $P_-$ & $P_i$ & $J_{ij}$ & $J_{+i}$ & $J_{+-}$ & $K_+$ & $\tilde{D}$ \\ \hline 
$P_+$ & $0$ & $0$ & $0$ & $0$ & $0$ & $-P_+$ & $0$ & $0$ \\ 
$P_-$ & $0$ & $0$ & $0$ & $0$ & $P_i$ &$ P_-$ & $-\tilde{D}$ & $2P_-$ \\ 
$P_i$ & $0$ & $0$ & $0$ & $P_j$ & $P_+$ & $0$ &$J_{+i}$ & $P_i$ \\ 
$J_{ij}$ & $0$ & $0$ & $-P_j$ & $J_{kl}$ & $J_{+i}$ & $0$ & $0$ &$0$ \\ 
$J_{+i}$ & $0$ & $-P_i$ & $-P_+$ & $-J_{+i}$ & $0$ &$-J_{+i}$ & $0$ & $-J_{+i}$ \\ 
$J_{+-}$ & $P_+$ & $-P_-$ & $0$ & $0$ & $J_{+i}$ &$0$ & $K_+$ & $0$ \\ 
$K_+$ & $0$ & $\tilde{D}$ & $-J_{+i}$ & $0$ & $0$ &$-K_+$ & $0$ & $-2K_+$ \\ 
$\tilde{D}$ & $0$ & $-2P_-$ & $-P_i$ & $0$ & $J_{+i}$ &$0$ & $2K_+$ & $0$ \\ \hline
  \end{tabular}
\caption{The commutation relation $i[X,Y]$ of very special conformal generators.}
\end{table}

\section{Local field theory examples}
Let us first recall the argument that very special conformal field theories with SIM(2) or HOM(2) invariance cannot be constructed from the spurion method.
Suppose we have a conformal field theory and try to deform it by adding local operators that preserve SIM(2) symmetry. In order to preserve the E(2) invariant very special conformal symmetry, which is a subgroup of SIM(2) invariant very special conformal symmetry, we try to add a vector primary operator
\begin{align}
 S = S_0 + \int d^4x \lambda^\mu J_\mu \ , \label{deform}
\end{align}
where $S_0$ is the action of a Poincar\'e conformal field theory, and  $\lambda^\mu$ has only non-zero components in $\lambda^+ $ so that it preserves $J_{+i}$ and $J_{ij}$.\footnote{More generically, we could add the tensor operators with only $+$ components, but the discussions below do not change.} In order to preserve the very special conformal symmetry $\tilde{D}$ and $K_+$, we further assume that the Poincar\'e scaling dimension of $J_\mu$ is five. This gives us a local field theory construction of E(2) invariant very special conformal field theory. The problem here is that the spurion vector $\lambda^\mu$ is not invariant under $J_{+-}$, and therefore, we cannot preserve the SIM(2) invariant very special conformal symmetry. This is essentially the reasoning made in \cite{Cohen:2006ky} to claim that there is no SIM(2) invariant (not necessarily conformal) field theories from the spurion method.\footnote{The argument does not rely on the conformal invariance, but note that it is based on the assumption that one can turn off the deformation such that the Lorentz invariance is recovered. This argument alone did not exclude the isolated examples if any.}

Nevertheless we do find a way to avoid this no-go argument by demanding that the spurion $\lambda^{\mu}$ transforms like a vector under $J_{ij}$ and $J_{+i}$, but transforms as a ``scalar" under $J_{+-}$. Since $J_{+-}$ does not appear in the right hand side of commutation relations of the very special conformal algebra, this causes no inconsistency at the level of the algebra. Of course, originally the spurion $\lambda^\mu$ was a vector under the full Lorentz transformation $J_{\mu\nu}$, so we need a trick to implement this idea.

The easiest way to do this is to use the concept of ``topological twist" \cite{Witten:1988ze}\cite{Witten:1988xj}. Suppose the original theory possesses an additional  non-compact global $U(1)$ symmetry $Q$. Suppose also that it has a vector operator $J_+$ which transforms as $ e^{-i\theta Q} J_+ e^{i\theta Q} = e^{\theta} J_+$ under the global symmetry $Q$. Then we see $\int dt d^3x J_+$ is invariant under $\tilde{J}_{+-} = J_{+-} + Q$ (while it was not invariant under $J_{+-})$. Now, we deform the action by the interaction
\begin{align}
 S = S_0 + \int d^4x \lambda^\mu J_\mu .
\end{align}
By construction, it is invariant under $J_{+i}$ and $J_{ij}$ as well as $\tilde{J}_{+-}$ as discussed above. The commutation relations among $J_{+i}$, $J_{ij}$ and $\tilde{J}_{+-}$ are the same as the ones in the very special relativity, so we may well regard $\tilde{J}_{+-}$ as $J_{+-}$ in the very special conformal algebra. In addition, if the Poincar\'e scaling dimension of $J_\mu$ is five, it preserves $\tilde{D}$ and $K_+$.
In this way, we have constructed a very special conformal field theory with the  SIM(2) invariance in a local fashion.\footnote{The similar idea to use the Poincar\'e dilatation rather than the global symmetry to twist $J_{+-}$ was discussed in \cite{Hariton:2006zj}. While the Lorentz part of the symmetry algebra is SIM(2), the commutator with the translation is different from the ones in table 1. 
We, therefore, called $\tilde{D}$ rather than $J_{+-}$ in our discussions.}

Let us show a couple of concrete examples to demonstrate the construction. First we consider a field theory with two real fields $A$ and $B$, which is defined by the action:
\begin{align}
S = \int d^3x dt \left(\partial_+ A \partial_- B + \partial_+ B \partial_-A - \partial_i A \partial_i B  + \lambda A^2 (A\partial_+ B - B \partial_+A) \right) \ . \label{AB}
\end{align}
Here the last term $\lambda A^2 (A\partial_+ B - B \partial_+A)$ plays the role of $\lambda^\mu J_\mu$ above.
It is obviously invariant under $P_+$, $P_-$ and $P_i$ as well as $J_{ij}$ and $J_{i+}$. It is invariant under the dilatation $\tilde{D}$:
\begin{align}
i[\tilde{D}, A(0)] &= A(0) \cr
i[\tilde{D}, B(0)] &= B(0)  
\end{align} 
as well as under the ``twisted" Lorentz boost $J_{+-}$:
\begin{align}
i[J_{+-}, A(0)] &= \frac{1}{2} A(0) \cr
i[J_{+-}, B(0)] &= -\frac{1}{2} B(0)  \ ,
\end{align}
where we omit the orbital part by setting $x^\mu =0$ because the invariance is trivial.
Since the deformation is given by a vector primary operator, it is invariant under a particular special conformal transformation $K_+$ 
\begin{align}
i[K_+, A(x)]  & = (2x^- + 2(x^-)^2 \partial_- + 2x^- x^i \partial_i + x_i^2 \partial_+) A(x) \cr
i[K_+, B(x)] &=  (2x^- + 2(x^-)^2 \partial_- + 2x^- x^i \partial_i + x_i^2 \partial_+) B(x) 
\end{align}
Therefore, this model is a concrete example of very special conformal field theories with the SIM(2) invariance.

Let us, however, mention one caveat of this model. The theory is non-unitary because of the wrong sign in the kinetic term. The underlying reason why we needed the non-unitarity is that we have to introduce the global non-compact $U(1)$ symmetry under which  real fields change their absolute values rather than the phases. This typically requires the kinetic term with the negative signature. In other words, it must be $SO(1,1)$ rather than $SO(2)$.

On the other hand, at the level of effective field theories, one may also construct a unitary field theory with the SIM(2) invariant very special conformal symmetry realized in a non-linear way. As an example, let us consider a field theory with a complex scalar $\phi$ and a real scalar $\varphi$ with the action
\begin{align}
 S &= \int d^3x dt \left( \partial_+ \phi^* \partial_- \phi + \partial_+ \phi \partial_-\phi^* - \partial_i \phi^* \partial_i \phi + |\phi|^2(\partial_+ \varphi \partial_- \varphi - \frac{1}{2}\partial_i \varphi \partial_i \varphi) \right. \cr
& \left. + i\lambda e^{\varphi}(\phi^* \partial_+ \phi - \phi \partial_+ \phi^* ) \right) \ .
\end{align}
To see how $J_{+-}$ symmetry is realized, we make $\varphi$ transform non-linearly under the dilatation $\tilde{D}$ and the Lorentz transformation $J_{+-}$
\begin{align}
i[\tilde{D},\phi] &= \phi \cr
i[\tilde{D},\varphi] &= 2 \cr
i[J_{+-},\phi] &= 0 \cr
i[J_{+-},\varphi] &= 1
\end{align}
so that the interaction $\int dt d^3x e^{\varphi}(\phi^* \partial_+ \phi - \phi \partial_+ \phi^* ) $ is invariant under $J_{+-}$ (as well as $\tilde{D}$). Note that the kinetic term is also invariant under the shift of $\varphi$.
While the action is invariant, this model breaks the dilatation and special conformal transformation spontaneously by choosing the vacuum expectation values of $\phi \neq 0$ to avoid the singular kinetic term for $\varphi$.

The similar construction is possible for the HOM(2) invariant case. Consider the action 
\begin{align}
 S &= \int d^3x dt \left( \partial_+ \phi^* \partial_- \phi + \partial_+ \phi \partial_-\phi^* - \partial_i \phi^* \partial_i \phi + |\phi|^2(\partial_+ \varphi\partial_- \varphi - \frac{1}{2}\partial_i \varphi \partial_i \varphi) \right. \cr
& \left. +i\lambda^{\mu\nu} e^{\varphi}(\partial_\mu \phi^* \partial_\nu \phi - \partial_\nu \phi^* \partial_\mu \phi ) \right) \ ,
\end{align}
where $\lambda^{\mu\nu} = -\lambda^{\nu\mu}$ has only non-zero component in $\lambda^{+2} = -\lambda^{2+}$.   We immediately see that the action is invariant under $P_\mu$ and $J_{+i}$ (but not under $J_{ij}$). Invariance under $J_{+-}$ is again guaranteed by the shift transformation of $\varphi$ field as $i[J_{+-},\varphi] = 1$ so that the interaction term $\int dt d^3x i\lambda^{\mu\nu} e^{\varphi}(\partial_\mu \phi^* \partial_\nu \phi - \partial_\nu \phi^* \partial_\mu \phi )$ becomes invariant. The theory is unitary, but it breaks the very special conformal symmetry spontaneously. The linear construction at the cost of unitarity is also possible with the action similar to \eqref{AB}.

In \cite{Nakayama:2017eof}, a holographic model for the E(2) invariant very special conformal field theory was discussed. Here, we, for the first time, present a holographic model for SIM(2) invariant very special conformal field theories.
Let us consider the five-dimensional Einstein gravity coupled with two real vector fields $A_M$ and $B_M$ with the action
\begin{align}
S = \int d^5x \sqrt{-g} \left(\frac{1}{2} R -\Lambda - \frac{1}{2}F_{MN}G^{MN} -  m^2 A_M B^M \right) \ ,
\end{align}
where $F_{MN} = \partial_M A_N - \partial_N A_M$ and $G_{MN} = \partial_M B_N - \partial_N B_m$. We set $\Lambda = -6$ and $m^2 = 8$. Then we find a particular solution of the equations of motion with the metric given by
\begin{align}
ds^2 = g_{MN} d^M dx^N = \frac{-2dx^+ dx^- + dx^i dx^i + dz^2}{z^2} \ 
\end{align}
and the vector fields:
\begin{align}
A = A_M dx^M = -\frac{dx^{-}}{z^2} \ , \ \  B = 0 \ . \label{vectors}
\end{align}
Invariance under $P_\mu$, $J_{+i}$, $J_{ij}$, $\tilde{D}$ and $K_+$ can be checked along the same line of discussions in \cite{Nakayama:2017eof}, where the holographic models for E(2) invariant very special conformal field theories are studied.
 Note, however, that in contrast with the model in \cite{Nakayama:2017eof}, the energy-momentum tensor  from the vector fields here is zero, so the geometry is not that of the Schr\"odinger holography \cite{Son:2008ye}\cite{Balasubramanian:2008dm} but it is just the AdS space-time.

Our claim is that this is a holographic dual description of a SIM(2) invariant very special conformal field theory. Naively, the vector condensation \eqref{vectors} is not invariant under the isometry of $J_{+-}$ while the metric is. Nevertheless, the crucial point is that the theory has a non-compact $U(1)$ global symmetry 
\begin{align}
\delta_{\lambda} A = e^{\lambda} A \ , \ \ \delta_{\lambda} B = e^{-\lambda} B \ ,
\end{align}
and the condensation becomes invariant under the new ``$J_{+-}$" if we define it  by a combined transformation of the coordinate transformation $dx^{-} \to e^{-\lambda}  dx^{-}$ and the non-compact global $U(1)$ symmetry $\delta_{\lambda}A = e^{\lambda} A$. This mechanism is essentially the holographic counterpart of what  we used in the field theory construction of conserved $J_{+-}$ from the idea of ``topological twist".\footnote{The ``topological twist" in the context of holography has been studied e.g. in \cite{Nakayama:2016ydc}\cite{Nakayama:2016xzs}.} Here the condensation of $A_M$ is equivalent to adding $J_\mu$ to the action. Similarly, the holographic theory is not unitary because of the wrong signs in the kinetic terms for the vector fields $A_M$ and $B_M$ much like the field theory construction discussed at the beginning of this section.

\section{Discussions} 
In this paper, we have constructed a local field theory example of SIM(2) or HOM(2) invariant very special conformal field theories, which was believed to be impossible within local quantum field theories. Our construction is either non-unitary or non-linear realization. We may regard these examples as counterexamples of no-go argument in \cite{Cohen:2006ky} with a little bit of a caveat. Now we are going to discuss what the caveat would imply.

The very special relativity was originally introduced from the motivations in elementary particle physics, but the existence of a particular null direction may have its origin from the other space-time physics. For example, let us imagine quantum field theories near a black hole (or black brane) horizon. There, the existence of a horizon may be associated with a particular null direction in space-time, and one may locally approximate the symmetry of the space-time by the very special relativity. We may even speculate that the difficulty of constructing SIM(2) or HOM(2) invariant field theories has its origin in black hole physics. On the one hand, we have to abandon locality to construct unitary theories. On the other hand, we have to abandon unitarity to construct local field theories. The locality vs unitarity in the black hole information puzzle has been a hot debate these days, and our discussions may be related to these studies in a deep manner.

We have also showed the local field theory construction of HOM(2) and SIM(2) very special conformal field theories with its non-linear realization by the spontaneous breaking. How does such non-linear realization appear in physics? We imagine that the very special relativity itself may be originated from the spontaneous symmetry breaking of the full Lorentz symmetry. In the black hole case above, this is what is precisely happening: the gravitational physics spontaneously breaks the Lorentz symmetry. Then we expect that the similar non-linear realization of the very special conformal symmetry may occur naturally.

Finally, beyond the spurion analysis in \cite{Cohen:2006ky}, there is no strict argument that very special conformal field theories with HOM(2) or SIM(2) invariance are impossible without violation of unitarity, violation of locality, or spontaneous breaking of the symmetry. It would be very important to prove or disprove this point. Such a no-go theorem (i.e. unitary Poincar\'e invariant field theories with $\tilde{D}$ and $K_+$ must be fully conformal invariant) does exist in $d=2$ dimensions \cite{Polchinski:1987dy}, and the analysis there suggests that we should understand the properties of correlation functions, in particular those of the energy-momentum tensor.\footnote{Some correlation functions in very special conformal field theories are currently studied by using the embedding method in \cite{Naka}.}


\section*{Acknowledgements}
This work is in part supported by JSPS KAKENHI Grant Number 17K14301.


\end{document}